\documentclass[12pt]{article}
\usepackage{epsfig}
\usepackage{relsize}
\usepackage{subfigure}
\usepackage[retainorgcmds]{IEEEtrantools}

\def\beq{\begin{equation}}
\def\eeq#1{\label{#1}\end{equation}}
\def\eeqn{\end{equation}}

\def\beqa{\begin{eqnarray}}
\def\eeqa#1{\label{#1}\end{eqnarray}}
\def\eeqan{\end{eqnarray}}

\let\bar=\overbar

\def\etal{{\it et al.}}

\def\Dslash{\not{\hbox{\kern-4pt $D$}}}
\def\dslash{\not{\hbox{\kern-2pt $\del$}}}

\def\BR{\mbox{\rm BR}}

\def\msb{{\bar{\ssstyle M \kern -1pt S}}}

\input babarsym.tex
\def\Title#1{\begin{center} {\Large {\bf #1} } \end{center}}
\def\Btonunub{\ensuremath{B^{0}\to} {\rm invisible}\xspace}
\def\Btonunubg{\ensuremath{B^{0}\to} {\rm invisible}\ensuremath{ + \gamma}\xspace}

\begin{document}
\begin{center}
Proceedings of CKM 2012, the 7th International Workshop on the CKM Unitarity Triangle, University of Cincinnati, USA, 28 September - 2 October 2012
\end{center}
\vspace{0.2cm}

\Title{Search for $B\to \nu\bar{\nu}$ and related modes with the \babar\ detector}

\bigskip\bigskip


\begin{raggedright}  
{\it Alessandro Rossi\index{Rossi, A.}, on behalf of \babar\ collaboration\\
Instituto Nazionale di Fisica Nucleare
- Sezione di Perugia\\
06123 Perugia, ITALY}
\bigskip
\end{raggedright}

\section{Introduction}
\begin{figure}[h]
  \centering
  \subfigure[]{
    \includegraphics[scale=0.6]{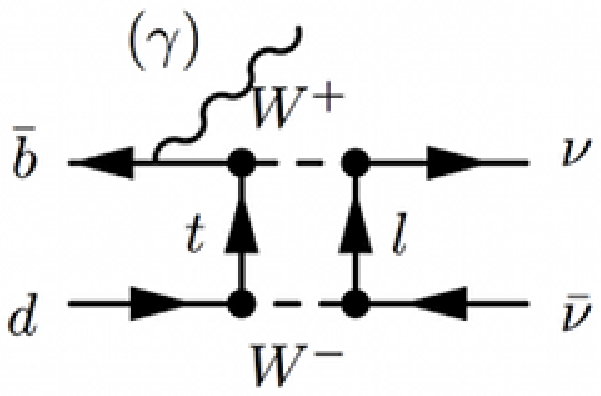}}
  \subfigure[]{
    \includegraphics[scale=0.6]{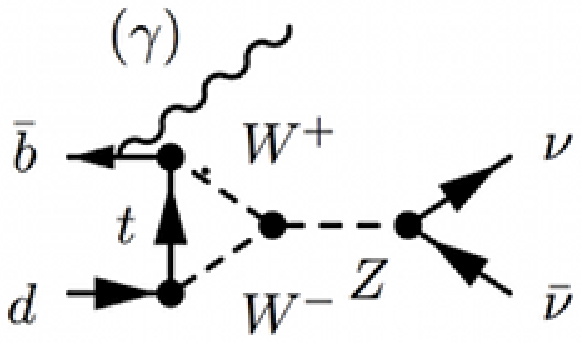}}
  \subfigure[]{
    \includegraphics[scale=0.6]{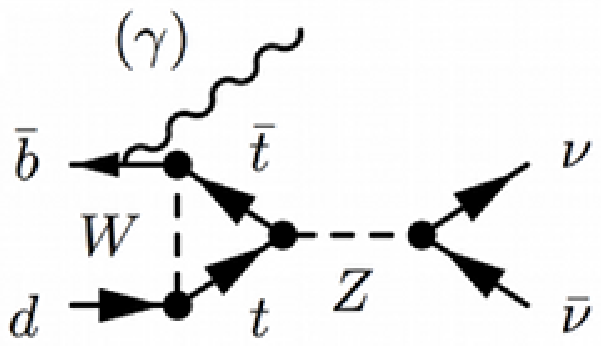}} 
    \label{fig:SMdecays}
    \caption{The lowest-order SM Feynman graphs for
 $B^{0}\to$ invisible (+ $\gamma$) decays: a) box diagram, b) and c) weak annihilation diagrams.}
\end{figure}

In the Standard Model (SM), a weak decay such as $B^{0}\to
\nu \bar{\nu} (+ \gamma$) can only occur through second-order diagrams like
those shown in Fig.~\ref{fig:SMdecays}.
All these processes are highly suppressed within the SM.
Like all purely leptonic $B$ decays, they contain a $b \to d$
transition plus an internal quark annihilation that further suppresses
the amplitude with respect to rare semileptonic decays. In addition,
helicity suppression factors proportional to $m_{\nu}^2$ make the
$\nu\bar{\nu}$ channel completely undetectable in the SM scenario.
For the $\nu\bar{\nu}\gamma$
channel the latter factor is not present, resulting in SM
branching fraction expectations at the $10^{-10}$
level~\cite{BNL787}.
Several new physics models predict enhancements of these branching ratios up to values close to the experimental detection: 
a phenomenological model allows for associated neutralino production from $B^0$ decays with a branching fraction in the $10^{-7}$ to $10^{-6}$ range~\cite{DDR}.  Also, models with large extra dimensions  can have the effect of producing significant, although small, rates for
invisible $B^0$ decays~\cite{ADW,AW,DLP}.

The data used in this analysis were collected with the \babar~detector at the \mbox{PEP-II} $e^{+}e^{-}$ collider at SLAC. The data sample corresponds to a luminosity of
424~fb$^{-1}$ accumulated at the $\Upsilon(4S)$ resonance and contains $(471 \pm 3)\times 10^{6}$ $B\bar{B}$ pair events.
For background studies we also used 45~fb$^{-1}$ collected at a center-of-mass (CM) energy about 40~MeV below the $B\bar{B}$ threshold (off peak).
A detailed description of the \babar~detector is presented in Ref.~\cite{BABARNIM}.

More details on these analysis may be found in Ref.~\cite{newprd}.

\section{Reconstruction and selection}

The detection of invisible $B$ decays uses the fact that
$B$ mesons are created in pairs, due to flavor conservation in $e^{+}e^{-}$ interactions.
We reconstruct events in which a $\bar{B}^0$ decays to $D^{(*)+}\ell^{-}\bar{\nu}$ (referred to as the ``tag side''), then look for consistency with an invisible decay or a decay to a single photon of the other neutral $B$ (referred to as the ``signal side'').

In the signal event selection we consider events with no charged tracks besides those from the $\bar{B}^{0} \to D^{(*)+}\ell^{-}\bar{\nu}$ candidate.
In order to reject background events where one charged or neutral particle is lost along the beam pipe, the cosine of the polar angle of the missing momentum in the CM frame ($\cos\theta^{*}_{miss}$) is required to lie in the $[-0.9,0.9]$ range.
\begin{figure}[!h]
\begin{center}
\includegraphics[width=8cm]{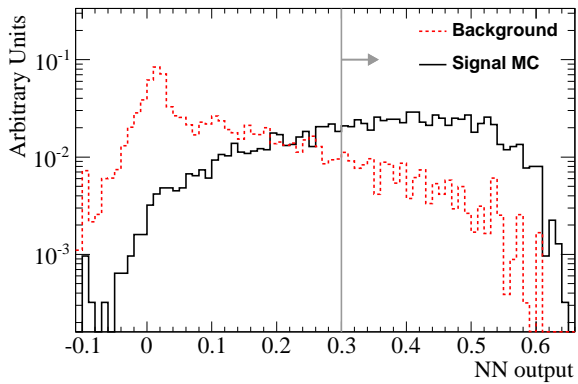}
\caption{Distributions of the NN output for simulated \Bz $\to$ invisible events with a $D$ meson on the tag side.
The black solid line is the signal while the red dashed line is the background. The solid gray vertical line
defines the NN output signal region.}
\label{fig:NNout}
\end{center}
\end{figure}
For the \Btonunub decay, in events where the $D$ meson on the tag side decays into $K^{-}\pi^{+}\pi^{-}$, two additional selection criteria are also applied:  on the sum of the angles between the Kaon and each one of the two $\pi$s and on the sum of the angles between the lepton and each one of the two $\pi$s.
To reconstruct \Btonunubg events, one remaining photon candidate with energy
greater than 1.2~\gev in the CM frame is also required.

An artificial neural network (NN) is used to provide further discrimination between signal and background events.
Events with a $D$ or a $D^{*}$ meson on the tag side are split in two different categories and a different NN was used for each sub-sample. For the \Btonunub decay 9 and 6 variables are used
for $D$ and $D^{*}$ sub-samples, respectively, while for the \Btonunubg decay we used 6 and 4 variables. These variables are mainly kinematical and refer to the tag side reconstruction. The two most important are the cosine of the angle between the $B$ meson and the $D^{(*)}\ell$ pair,  defined as
\begin{equation}
\hspace*{-0.2cm}\cos \theta_{B,D^{(*)-}\ell^{+}} = \frac{2\,E_{B} E_{{D^{(*)-}\ell^+}} -m^2_{{\B}} - m^2_{{D^{(*)-}\ell^+}}}
{2\,|\vec{p}_{B}| |\vec{p}_{{D^{(*)-}\ell^+}}|   },
\end{equation}
and $M_{miss}^{tag}$ (defined as the invariant mass of the event after the $D^{(*)-}\ell^{+}$ pair is subtracted). In the \Btonunubg analysis, we additionally use the energy of the photon on the signal side, evaluated in the laboratory frame. In Fig.~\ref{fig:NNout}, the output of the NN for simulated \Btonunub with a $D$ meson on the tag side,
and the corresponding signal region, are shown.

After the NN selection, the $D$ meson invariant mass ($m_{D}$) and the difference between the reconstructed $D^{*}$ invariant mass and the PDG \Dz mass ($\Delta m$) are used to define a 
\begin{figure}[!h]
\hspace{1.6cm}
\includegraphics[width=6cm]{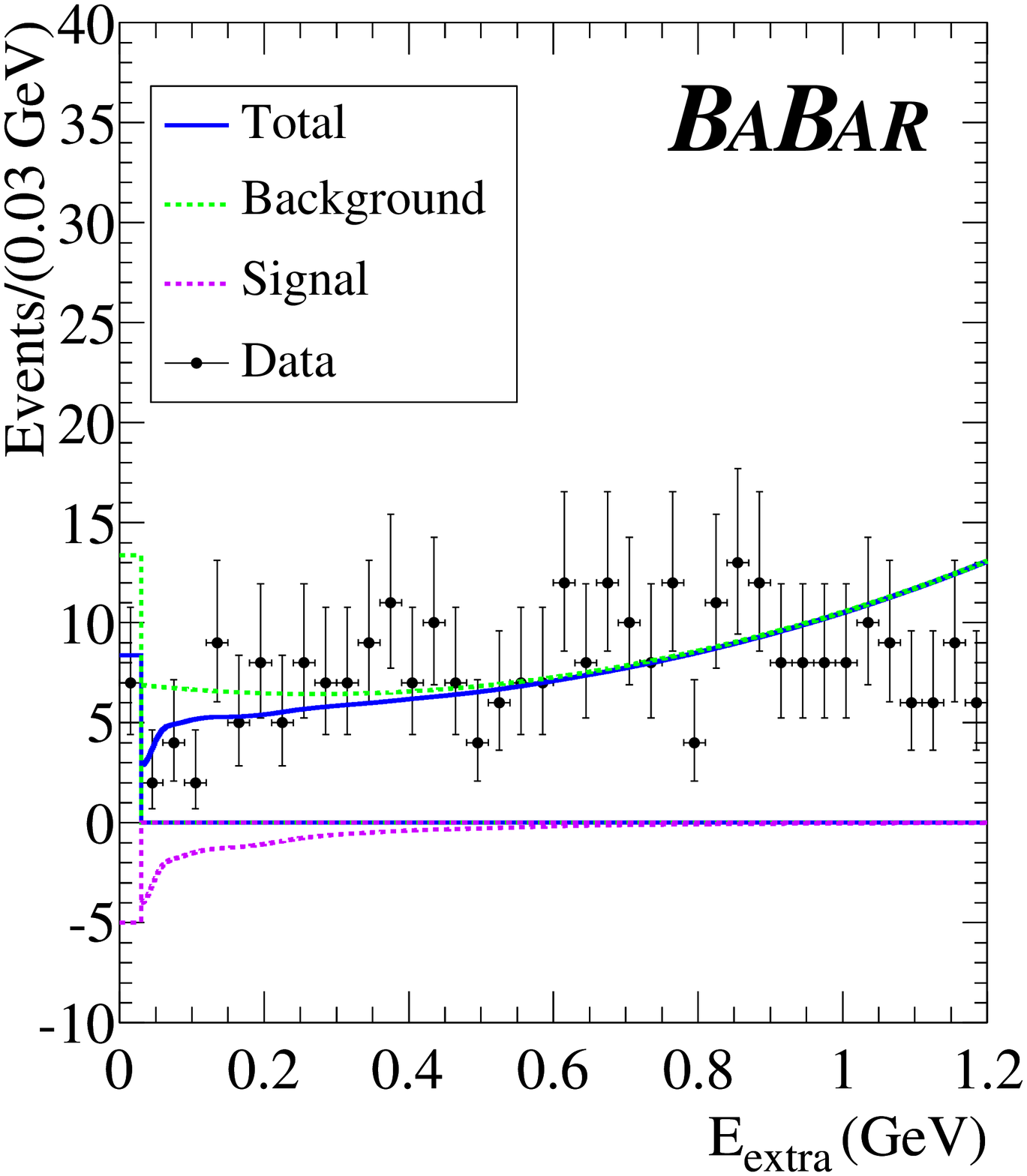}
\includegraphics[width=6cm]{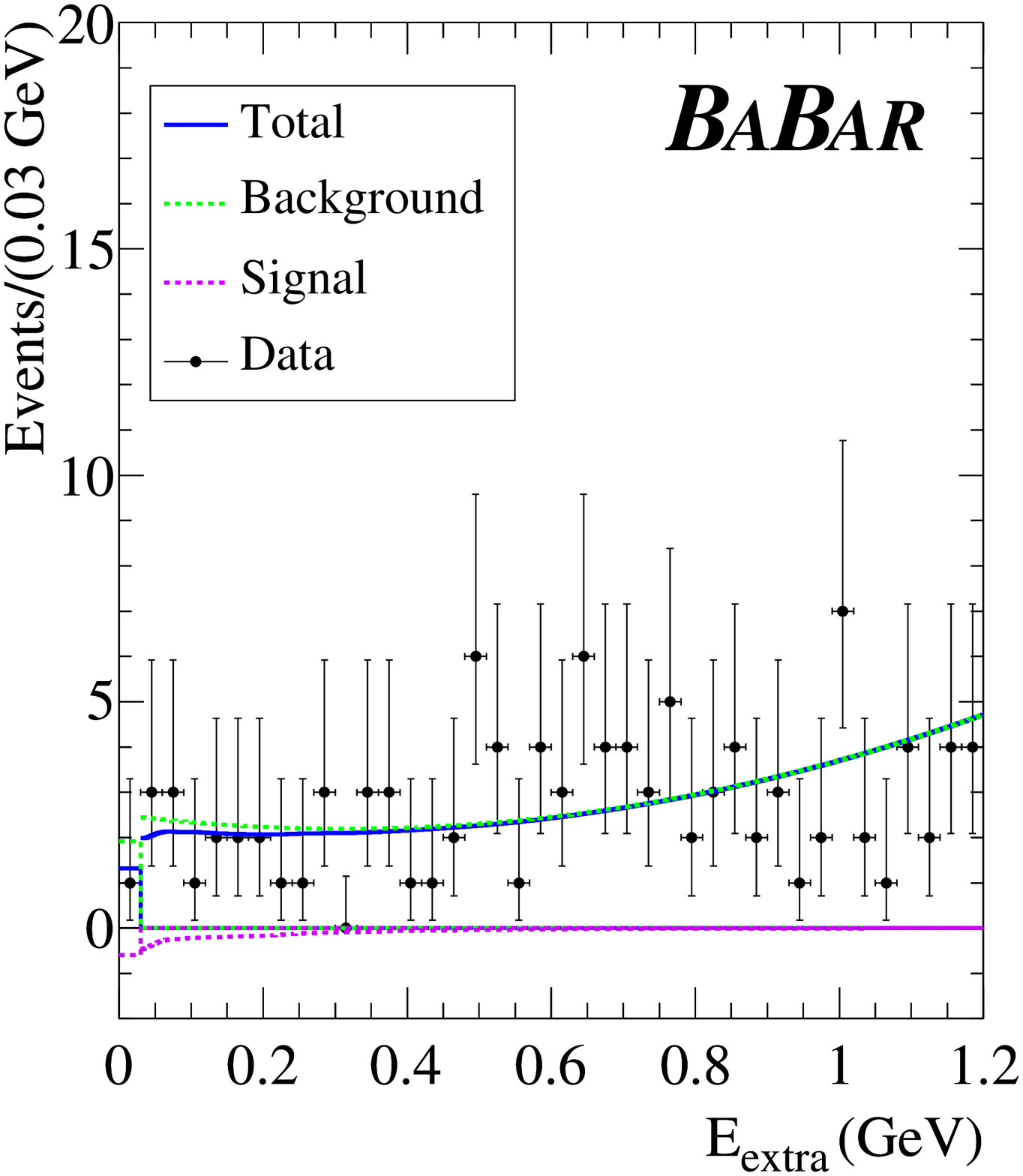}
\caption{Results of the maximum likelihood fit of $E_{\rm extra}$ for \Btonunub (left) and \Btonunubg (right).}
\label{fig:etotleft}
\end{figure}
signal region and a sideband region for the $D$ tag and \Dstar tag samples, respectively.
The signal region is defined as a $\pm$15~\mevcc window around the PDG value for $m_{D}$, and as 0.139 $< \Delta m <$ 0.148~\gevcc. The excluded regions are used as sidebands.

The neutral energy that remains after all tag side tracks and neutral clusters have been accounted for is denoted as  $E_{\rm extra}$.  For \Btonunubg, the energy of the highest-energy photon remaining in the event (the signal photon candidate) is also removed from the $E_{\rm extra}$ computation. The $E_{\rm extra}$ signal region is defined by imposing an upper bound at 1.2~\gev.

We construct probability density functions (PDFs) for the $E_{\rm extra}$ distribution
for signal ($\mathcal{P}_{\rm sig}$) and background ($\mathcal{P}_{\rm bkg}$) using
detailed MC simulations for signal and data from the $m_{D}$ and $\Delta m$ sidebands for background.
The two PDFs are combined into an extended maximum likelihood function
$\mathcal{L}$, defined as a function of the free parameters $N_{\rm sig}$ and $N_{\rm bkg}$,
the number of signal and background events, respectively.
The photon reconstruction algorithm has a lower cluster energy cut of
  30~\mev, and as a consequence, the $E_{\rm extra}$ distribution is not continuous.
To account for this effect, the likelihood is composed of two distinct parts, one for $E_{\rm extra}>30$~\mev and one for $E_{\rm extra}=0$~\mev. 

The negative log-likelihood
is then minimized with respect to
$N_{\rm sig}$ and $N_{\rm bkg}$ in the data sample.
The resulting fitted values for $N_{\rm sig}$ and $N_{\rm bkg}$ are given in Table~\ref{yields}.
Figure~\ref{fig:etotleft} shows the $E_{\rm extra}$ distributions for
\Btonunub and \Btonunubg with the fit superimposed.

\begin{table}[hb]
  \begin{center}
    \begin{tabular}{l|cc}
      Mode       & $N_{\rm sig}$ & $N_{\rm bkg}$ \\
      \hline
     \Btonunub  & $-22 \pm 9$    & $334 \pm 21$  \\
     \Btonunubg & $-3.1 \pm 5.2$ & $113 \pm 12$  \\
      \hline
    \end{tabular}
  \end{center}
    \caption{   \label{yields}Fitted yields of signal and background events in data.  The uncertainties are statistical.}
\end{table}

Using detailed Monte Carlo simulations of \Btonunub and \Btonunubg events,
we determine our signal efficiency to be $(17.8 \pm 0.2) \times 10^{-4}$
for \Btonunub and $(16.0 \pm 0.2) \times 10^{-4}$ for \Btonunubg, where the uncertainties
are statistical.
\begin{table}[ht]
\resizebox{\textwidth}{!}{
    \begin{tabular}{l|cc|l|cc}
      Source & \Btonunub & \Btonunubg & Source & \Btonunub & \Btonunubg\\
      \hline
      \multicolumn{3}{c|}{\bf Normalization Errors} & \multicolumn{3}{c}{\bf Efficiency Errors}\\
      $B$-counting & $0.6\%$ & $0.6\%$ & Tagging Efficiency & $3.5\%$ & $3.5\%$\\
      \cline{1-3} 
      \multicolumn{3}{c|}{\bf Yield Errors (events)} &$m_{D}$ $(\Delta m)$ Selection & $1\%$ & $1.3\%$\\
      Background Param. & $15.8$ & $6.5$ &  Preselection   & $3\%$ & $2.4\%$\\
      Signal Param. & $2.0$ & $1.2$ & Neural Network & $6.1\%$ & $8.2\%$\\
      Fit Technique  & -- & $1.0$ & Single Photon & -- & $1.8\%$\\
      \cline{4-6}
      $E_{\rm extra}$ Shape & $0.1$ & $1.8$ & TOTAL & $7.7\%$ & $9.5\%$\\
      \hline
      TOTAL & $15.9$ & $6.9$ & \multicolumn{3}{c}{}\\
      \cline{1-3}
    \end{tabular}
}    
    \caption{\label{totsys}Summary of the systematic uncertainties.}
\end{table}

The systematic uncertainty on the signal efficiency is dominated
by data-MC discrepancies in the distribution of the variables used as input to the NN while the main 
systematic uncertainty on the signal yield is dominated by the background parametrization uncertainties.
The total systematic uncertainty on the signal selection efficiency is 7.7\% for \Btonunub decay and 9.5\% for \Btonunubg decay and the total systematic errors on the signal yield are 16 and 7 events for \Btonunub and \Btonunubg, respectively. All systematic uncertainties are summarized in Table~\ref{totsys}.

\section{Conclusions}
A Bayesian approach is used to set 90\% confidence level (CL) upper limits on the branching fractions for \Btonunub and \Btonunubg.
Flat prior probabilities are assumed for positive values of both branching fractions.  Gaussian likelihoods are adopted for signal yields.
The Gaussian widths are fixed to the sum in quadrature of the statistical and systematic yield errors. We extract a posterior PDF using Bayes' theorem,
including in the calculation the effect of systematic uncertainties associated with the efficiencies and the normalizations, modeled by Gaussian PDFs.
Given the observed yields in Table~\ref{yields}, the 90\% CL upper limits are 
\begin{eqnarray}
\mathcal{B}(\Btonunub) & < & 2.4 \times 10^{-5}\nonumber\\
\mathcal{B}(\Btonunubg) & < & 1.7 \times 10^{-5}\nonumber
\end{eqnarray}
at 90\% CL.
These limits supercede our earlier results~\cite{prevprl}, which used a small fraction of our present dataset, and the recent Belle result \cite{belle}.

\def\Discussion{
\setlength{\parskip}{0.3cm}\setlength{\parindent}{0.0cm}
     \bigskip\bigskip      {\Large {\bf Discussion}} \bigskip}
\def\speaker#1{{\bf #1:}\ }
\def\endDiscussion{}


\begin{thebibliography}{99}

\bibitem{BNL787} 
S.~Adler \etal\ (BNL-787 Collaboration), \textit{Phys. Rev. Lett.} 84, 3768 (2007);

\bibitem{DDR}
A.~Dedes, H.~Dreiner, and P.~Richardson,
\textit{Phys. Rev. D} 65, 015001 (2002).                                                                                              

\bibitem{ADW}
K.~Agashe, N.~G.~Deshpande, and G.-H.~Wu,
\textit{Phys. Lett. B} 489, 367 (2000).

\bibitem{AW}
K.~Agashe and G.-H.~Wu,
\textit{Phys. Lett. B} 498, 230 (2001).

\bibitem{DLP}
H.~Davoudiasl, P.~Langacker, and M.~Perelstein,
\textit{Phys. Rev. D} 65, 105015 (2002).

\bibitem{BABARNIM}
B.~Aubert {\it et al.}  (\babar~Collaboration),
\textit{Nucl.\ Instrum.\ Meth.\ A} 479 (2002).

\bibitem{newprd}
J.~P.~Lees {\it et al.}  (\babar~Collaboration),
\textit{Phys.\ Rev.\ D} 86, 051105 (2012).

\bibitem{prevprl}
B.~Aubert \etal\ (\babar Collaboration),  \textit{Phys. Rev. Lett.} 93, 091802  (2004);

\bibitem{belle}
C.~L.~Hsu {\it et al.}  [Belle Collaboration],
  \textit{Phys.\ Rev.\ D}  86, 032002 (2012).

\end{thebibliography}
\end{document}